# Multiplicity distributions of the charged particles at the maximum of electromagnetic showers initiated by 5-1000 GeV electrons in Fe, W and Pb

S. P. Denisov, V. N. Goryachev

**Institute for High Energy Physics  
of the National Research Centre  
«Kurchatov Institute»**

*e-mail: denisov@ihep.ru*

**Abstract**

**Charged particles multiplicity distributions at the maximum of electromagnetic showers initiated by 5 to 1000 GeV electrons in Fe, W, and Pb were calculated using GEANT4. It is shown that they are reasonably well fitted by the inverse sum of two exponents and the energy dependence of the average multiplicity follows power law with the power of ~0.95 for all studied materials.**

# 1.Introduction

Detectors consisting of a high $Z$ converter and a hodoscope type particle detector behind it are often used in HEP experiments for $e,\gamma$ /hadron and $\gamma/\pi^0$ separations and for $e,\gamma$ coordinate and energy measurements[1-18]. The most popular converter materials are Pb and W, while Fe or Cu are used less frequently. The converter thickness $t$ usually varies from 2-3 $X_0$ to $t_{max}$ where $X_0$ is a radiation length and $t_{max}$ corresponds to the maximum flux $N_{max}$ of charge particles in the electromagnetic (EM) shower. A converter of $t_{max}$ placed in high energy electron beam can also be used as a source of the short and intense bunches of relativistic positrons and electrons[19]. Thus the characteristics of EM showers at $t_{max}$ are of particular interest. The charged particles flux at $t_{max}$ consists mainly of $e^+$ and $e^-$. For example for 200 GeV electrons hitting Pb it contains 56% of $e^-$, 44% of $e^+$ and 0.018% of other particles.

In this paper the results of calculations of the charged particles multiplicity distributions at $t_{max}$ for the Fe, W and Pb converters irradiated by 5 to 1000 GeV electrons are presented. All converters are 70 cm in diameter. The calculations are based on GEANT4 10.01.p02 (Physical list FTFP_BERT)[20, 21]. By default the range cut of $R_c$=0.7 mm is used for the shower particles in this version. Corresponding energy thresholds are shown in Table 1. They are much less than the average $e^-$ and $e^+$ energy of ~50 MeV at $t_{max}$. Thus one can expect rather weak $N_{max}(R_c)$ dependence. Calculations performed for twice less and twice higher $R_c$ values confirm this conclusion (see Table 2). All results presented below were obtained with $R_c$=0.7 mm.

Table 1. Energy thresholds for $R_c$=0.7 mm[20] and radiation lengths[22].

| Material | Energy thresholds, MeV | | | | $X_0$, g/cm$^2$ |
|---|---|---|---|---|---|
| | gamma | electron | positron | proton | |
| Fe | 0.017 | 0.951 | 0.902 | 0.070 | 13.84 |
| W | 0.097 | 1.640 | 1.543 | 0.070 | 6.76 |
| Pb | 0.095 | 1.004 | 0.951 | 0.070 | 6.37 |

Table 2. $N_{max}(R_c)$ in Pb for 20 and 200 GeV electrons.

| Cut, mm | 0.35 | 0.7 | 1.4 |
|---|---|---|---|
| 20 GeV | 96.4 ± 0.6 | 95.2 ± 0.6 | 94.1 ± 0.6 |
| 200 GeV | 828.1 ± 3.5 | 825.9 ± 3.5 | 825.6 ± 3.5 |

## 2. Energy dependence of $t_{max}$ and $N_{max}$

To find the converter thicknesses $t_{max}$ from 500 to 4000 EM cascades were generated for the primary electron energies $E_0$ of 5, 10, 20, 30, 40, 80, 120, 160, 200, 300, 500 and 1000 GeV. Obtained dependencies of the average number $<N>$ of charged particles *vs* converter depth $t$ are shown in Fig.1. They are fitted by gamma function

$$<N> = c_0 (bt)^{a-1} e^{-bt} \qquad (1)$$

where $c_0$, $a$ and $b$ are free parameters [23]. Function (1) reaches maximum at

$$t_{max} = (a-1)/b. \qquad (2)$$

$a$ and $b$ values obtained by fitting are presented in Fig. 2. As expected[22] $a$ depends logarithmically on $E_0$:

$$a = a_1 \ln E_0 + a_2, \qquad (3)$$

where $a_{1,2}$ are free parameters shown in Table 3. It is evident from Fig. 2 that parameter $b$ is independent of energy and its averaged values are equal to 0.580 (Fe), 0.537(W) and 0.535(Pb) in $1/X_0$ units. One need to keep in mind that $a$ and $b$ are correlated [24].

The values of $t_{max}$ and $N_{max}=<N>(t=t_{max})$ calculated by formulas (1) and (2) are shown in Fig. 3. $t_{max}$ follows the same $E_0$ dependence as $a$:

$$t_{max} = c_1 \ln E_0 + c_2, \qquad (4)$$

where $c_{1,2}$ are free parameters shown in Table 3. $N_{max}$ as a function of $E_0$ follows power law:

$$N_{max} = N_0 E_0^k \qquad (5)$$

with power $k$ close to 0.95 (Table 3) in agreement with previous calculations[2] and experimental results[4]. As can be seen from Figs. 2, 3 and Table 3, all values for Pb and W parameters are close to each other and charge particle flux in Fe is by factor of 1.4 less than that in W and Pb in agreement with measurements[25].

Table 3. Parameters values in the equations (3)-(5), $E_0$ in GeV, $c_{1,2}$ in $X_0$.

| Material | Fe | W | Pb |
|---|---|---|---|
| $a_1$ | 0.586±0.012 | 0.621±0.019 | 0.599±0.015 |
| $a_2$ | 2.60±0.06 | 2.57±0.09 | 2.66±0.07 |
| $c_1$ | 1.05±0.01 | 1.13±0.01 | 1.11±0.01 |
| $c_2$ | 2.61±0.03 | 3.06±0.06 | 3.14±0.05 |
| $N_0$ | 3.92±0.03 | 5.44±0.02 | 5.39±0.03 |
| $k$ | 0.946±0.001 | 0.945±0.001 | 0.950±0.001 |

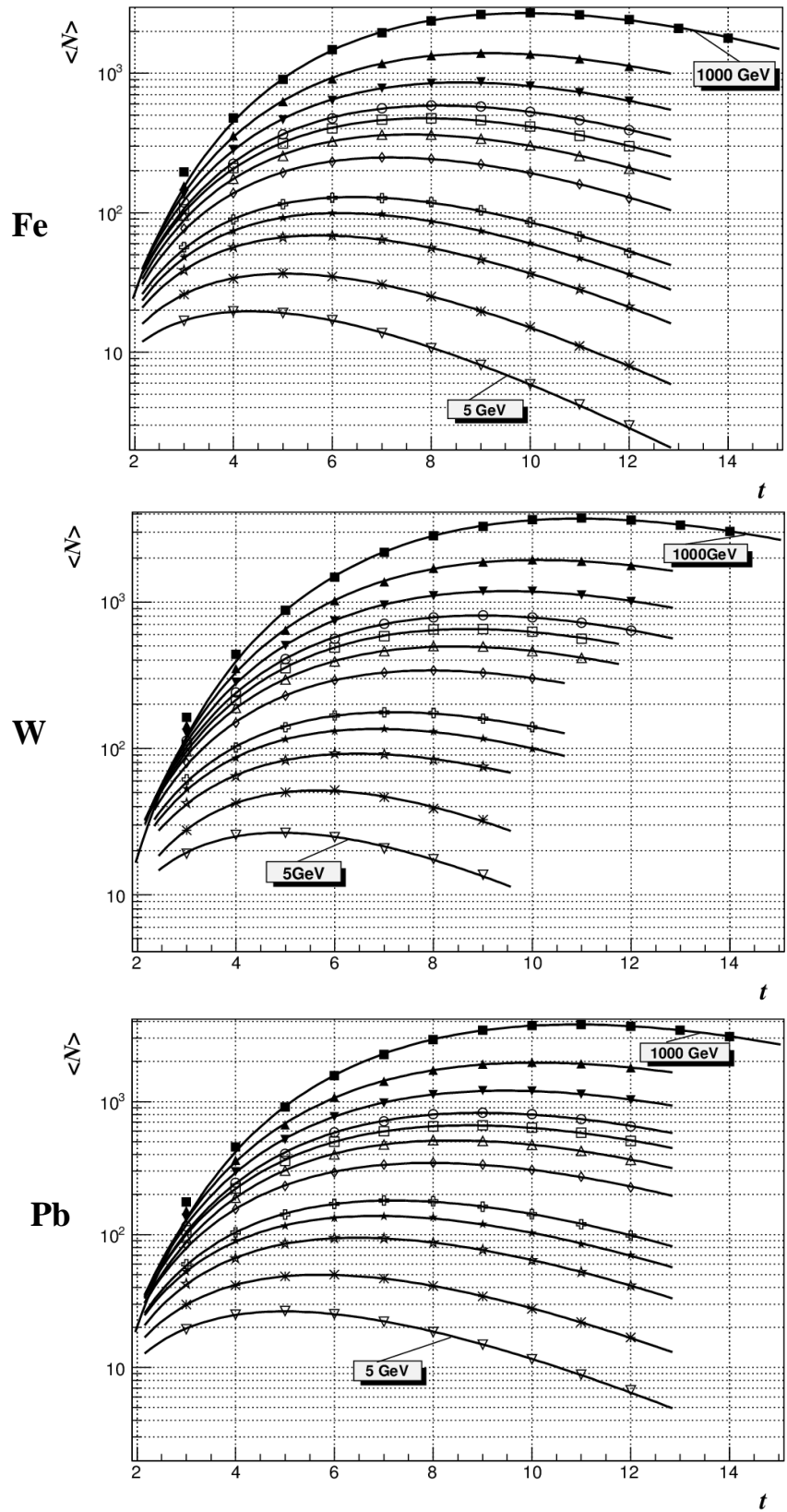

Fig. 1. Results of GEANT4 simulation of EM showers fitted by gamma function (1).

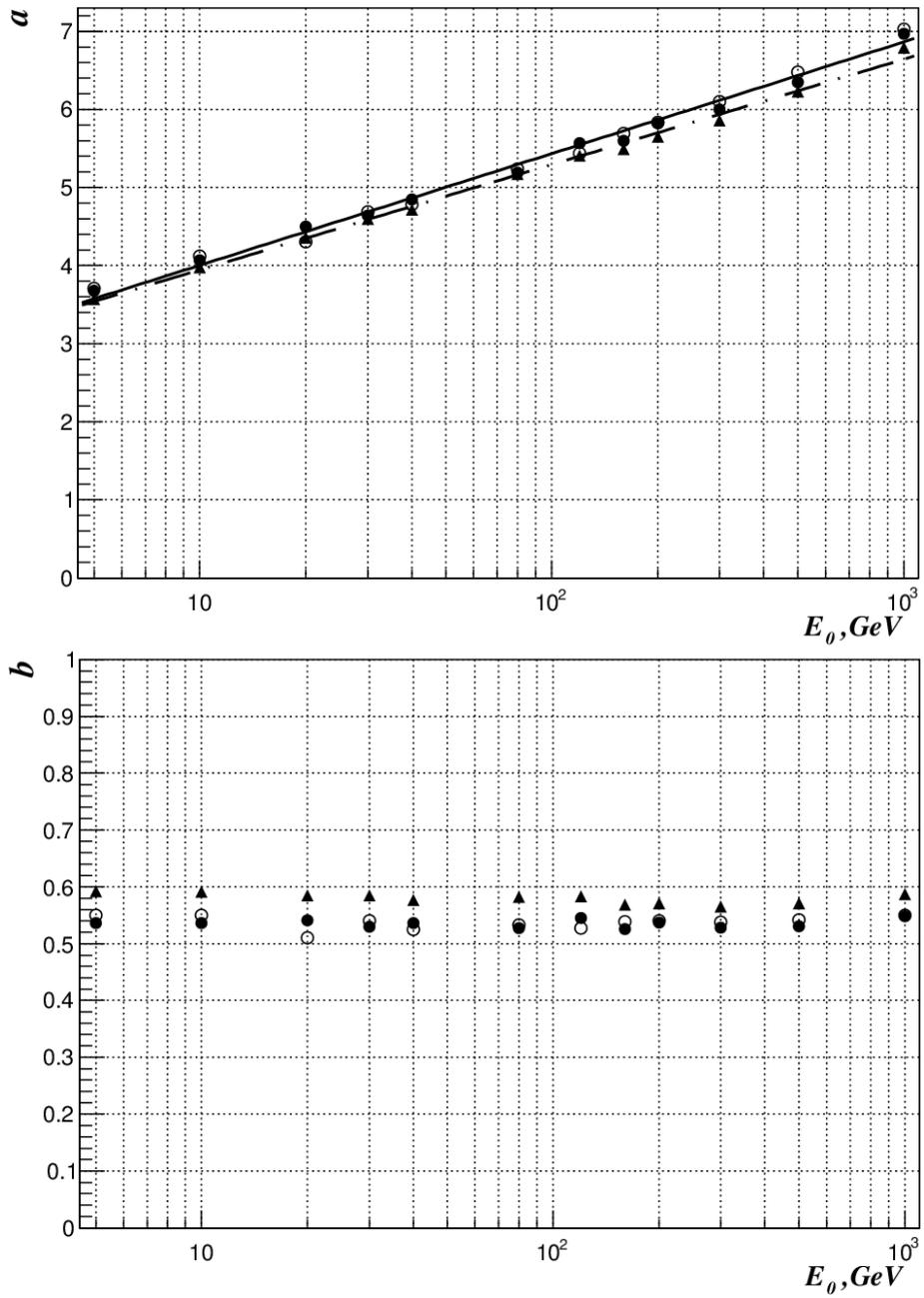

Fig. 2. *a* and *b* values *vs* $E_0$ for Fe (▲), W (○) and Pb (●). Parameter *b* is in $1/X_0$ units. The solid and dash-dotted lines are the fits to formula (3) for W and Fe with $a_1$ and $a_2$ parameters shown in Table 3. Results for W and Pb almost coincide.

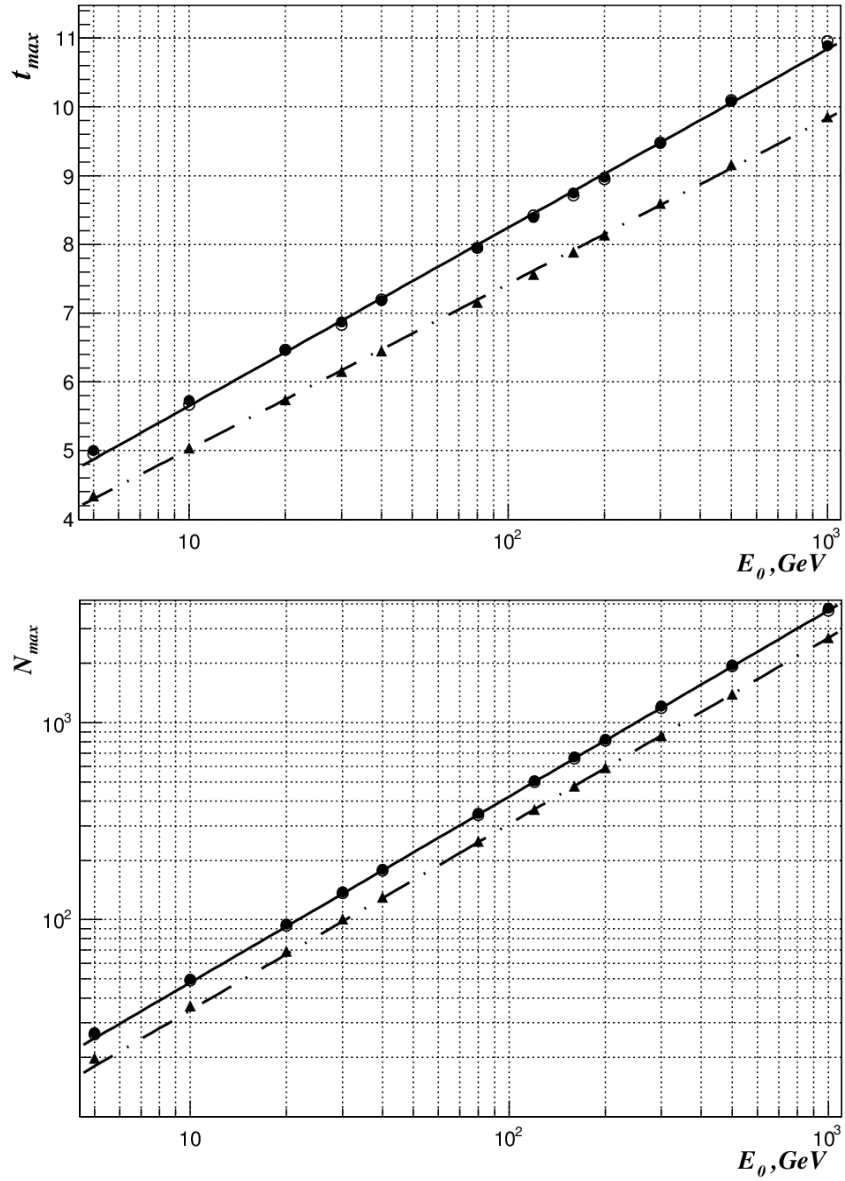

Fig. 3. $t_{max}$ (in $X_0$ units) and $N_{max}$ vs $E_0$ for Fe (▲), W (○) and Pb (●). The solid (W) and dash-dotted (Fe) lines are the fits to formulas (4) and (5) with parameters shown in Table 3. Results for W and Pb almost coincide.

## 3. Multiplicity distributions

Calculated probability distributions $dP/dN$ of charged particles multiplicity at $t_{max}$ are shown in Figs. 4-9. They have tails at low multiplicities due to late development of some showers. Several functions representing different combinations of exponents, Gaussians and polynomials were used to describe these distributions. The best fits for the entire range of $E_0$ were obtained for the inverse sum of two exponents:

$$\frac{dP}{dN} = \frac{p_0}{e^{p_1(N-p_3)} + e^{p_2(N-p_3)}}, \qquad (6)$$

where $p_0$ is a normalization factor and $p_1$, $p_2$, $p_3$ are free parameters. The function (6) is defined from $-\infty$ to $+\infty$. Its integral is equal to

$$\int_{-\infty}^{\infty} \frac{dP}{dN} dN = \int_{-\infty}^{\infty} \frac{p_0 e^{-pN'}}{1+e^{qN'}} dN' = p_0 \frac{\pi}{q} \operatorname{cosec} \frac{p\pi}{q},$$

if $q>p>0$ or $0>p>q$ where $p=p_1$ и $q=p_1-p_2$[26]. Thus the factor $p_0$ can be presented in the following form:

$$p_0 = \frac{p_1 - p_2}{\pi} \sin \frac{\pi p_1}{p_1 - p_2} \qquad (7)$$

Parameters $p_1$ and $p_2$ define the steepness of the right and left slopes of $dP/dN$ and $p_3$ is close to the most probable value

$$N_{mp} = p_3 - \frac{\ln p_1 - \ln(-p_2)}{p_1 - p_2} \qquad (8)$$

(the second term in (8) is ~3% of $p_3$ for all materials and energies).

The results of the fit of $dP/dN$ distribution using formula (6) are presented in Figs. 4-10 and Tables 4-7. Fitting is performed for the entire range of $N$ from 0 to the maximum value shown in the corresponding figure. It turned out that $p_3$ follows the same $E_0$-dependence as $N_{max}$ (5):

$$p_3 = P_0 E_0^l \qquad (9)$$

Values of the free parameters $P_0$ and $l$ are presented in the Table 4. From Tables 3 and 4 it follows that powers $l$ and $k$ are close to each other.

Table 4. Parameters in formula (9), $E_0$ in GeV.

| Convertor | Fe | W | Pb |
| --- | --- | --- | --- |
| $P_0$ | 4.321±0.049 | 5.726±0.091 | 6.094±0.027 |
| $l$ | 0.945±0.002 | 0.951± 0.002 | 0.943±0.001 |

Formula (6) can't be used at small values of $<N>$ when $dP/dN$ (0) is not close to 0.

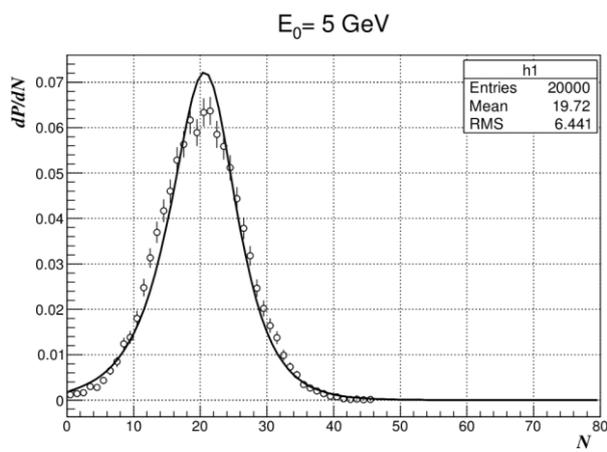
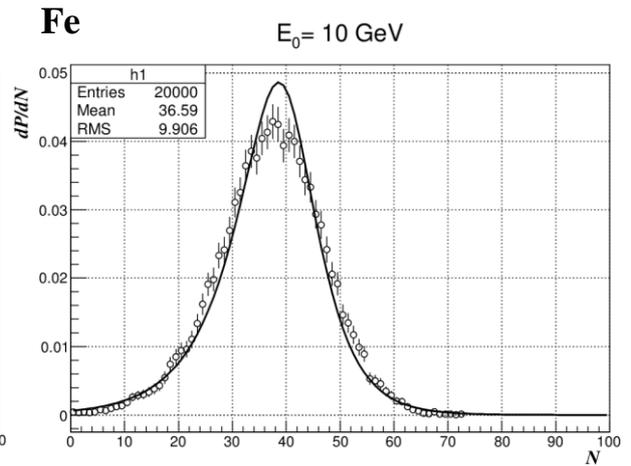

**Fe**

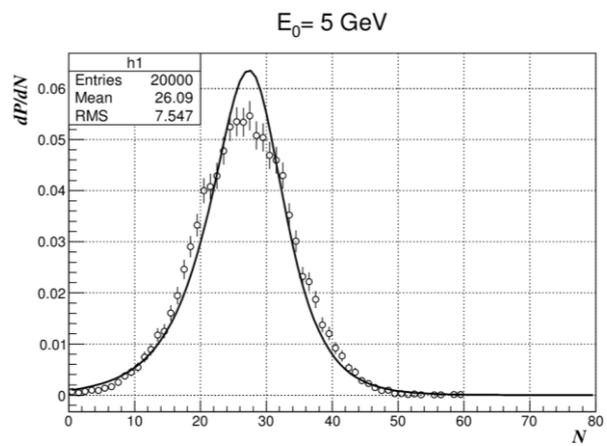
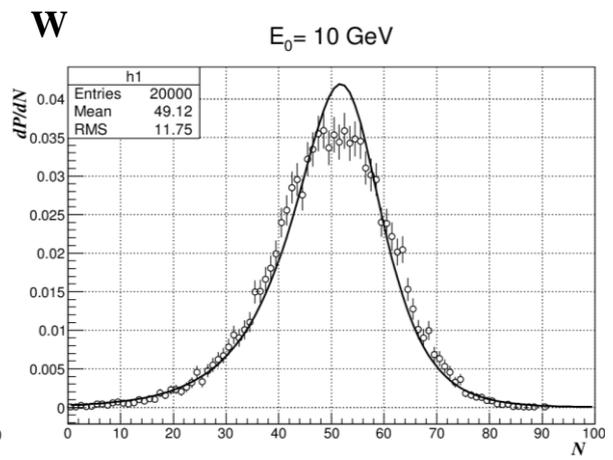

**W**

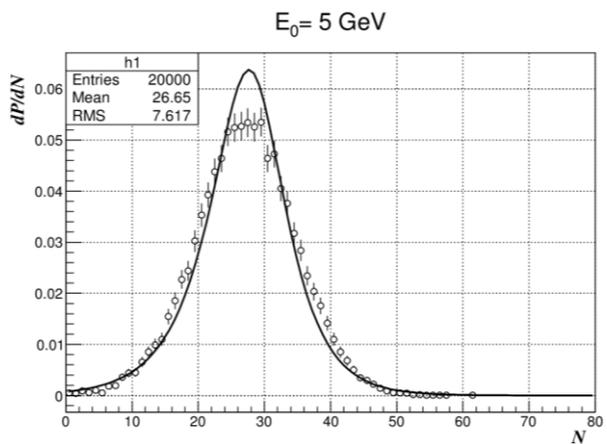
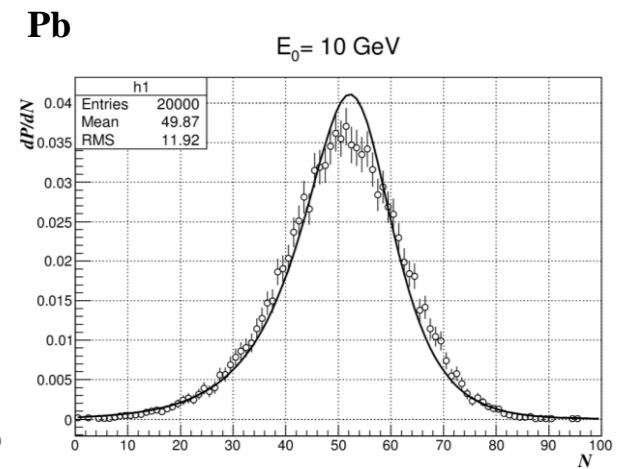

**Pb**

Fig. 4. Multiplicity distributions for the shower energies of 5 and 10 GeV fitted to formula (6).

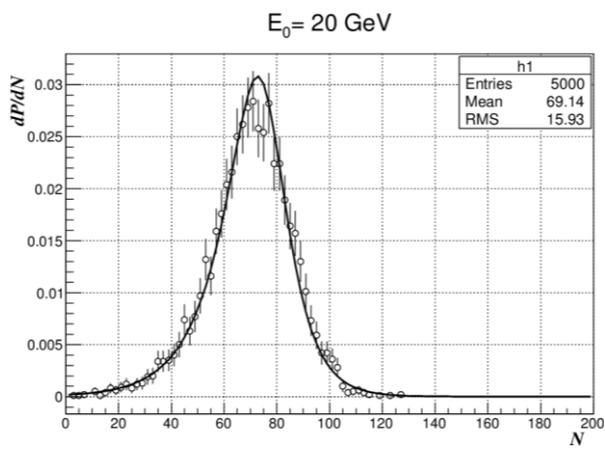
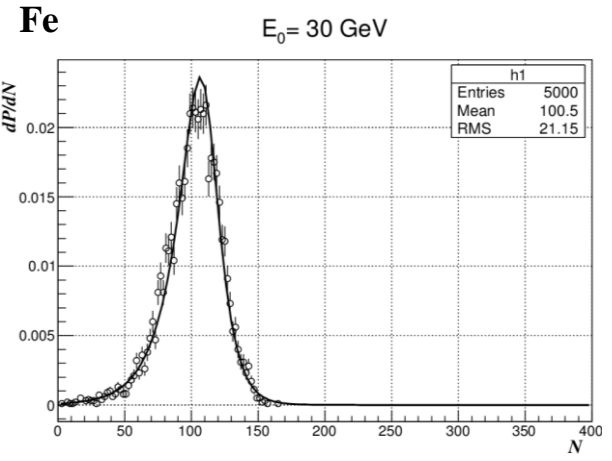
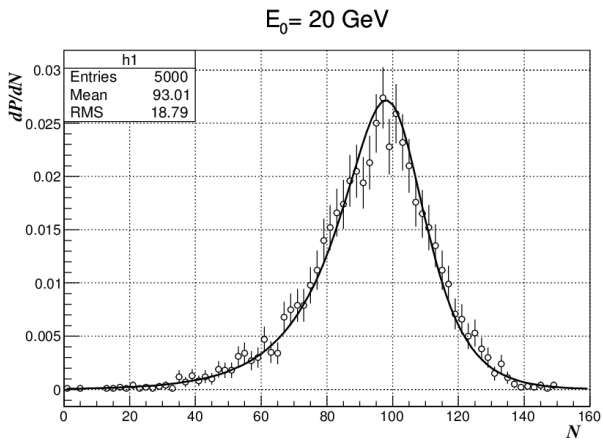
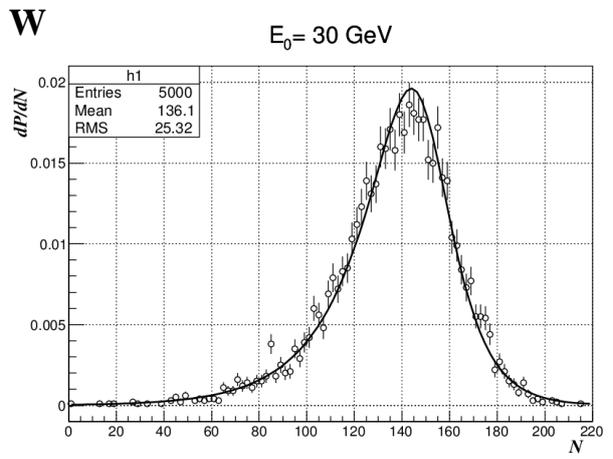
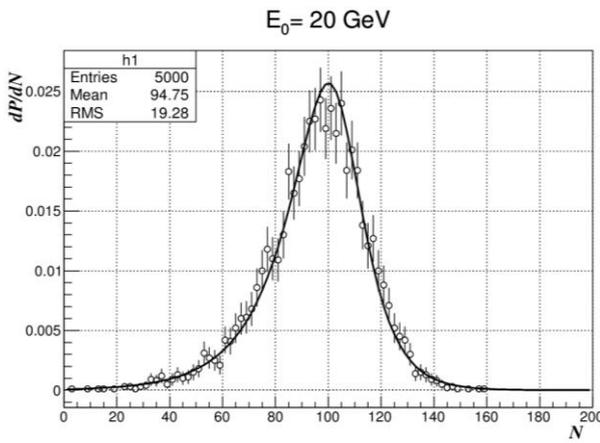
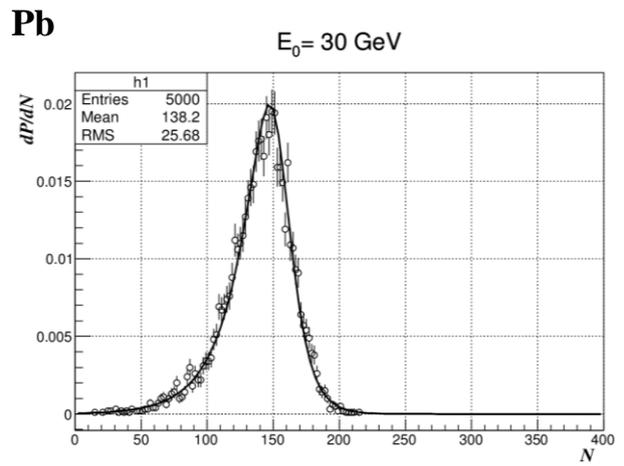

Fig. 5. Multiplicity distributions for the shower energies of 20 and 30 GeV fitted to formula (6).

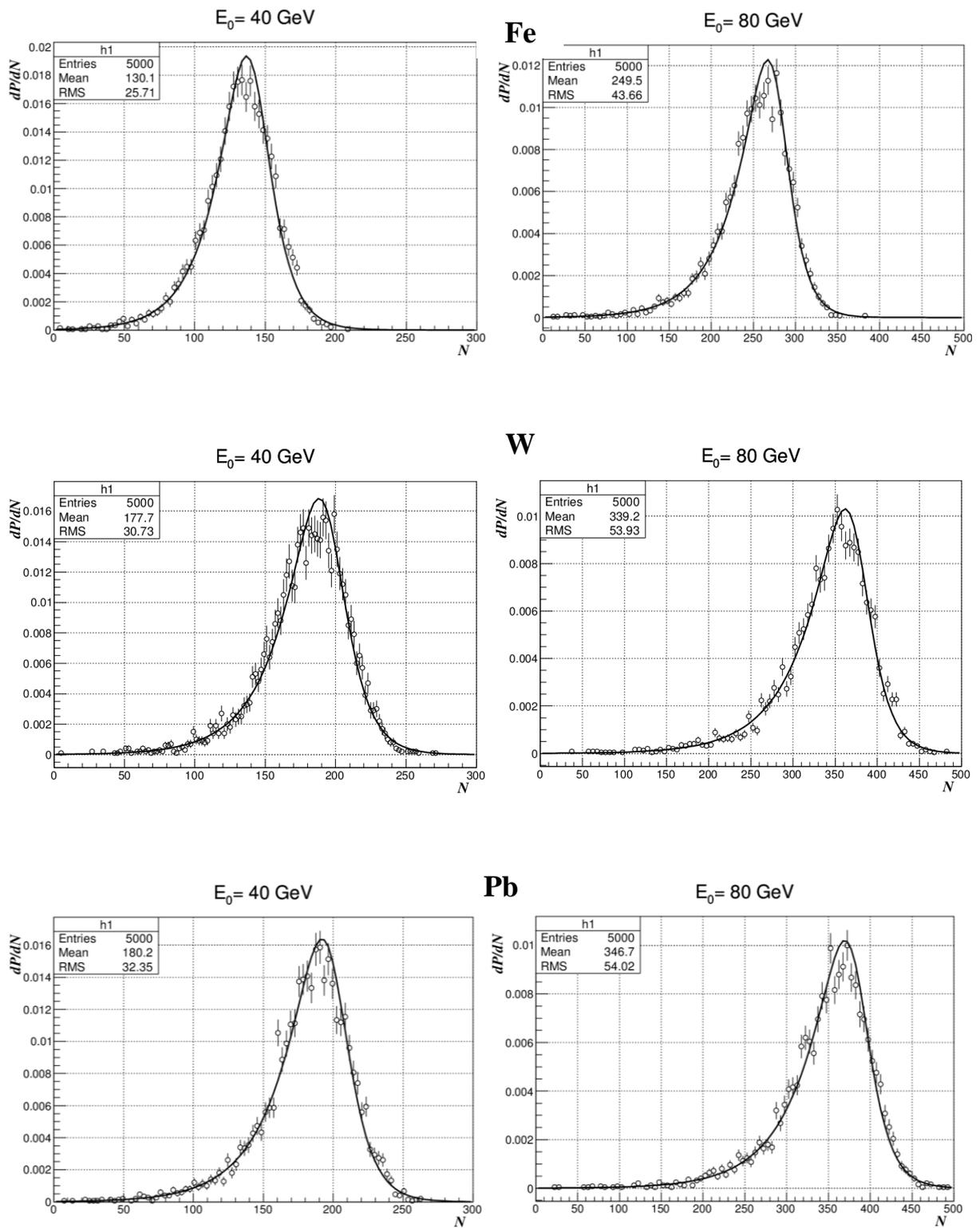

Fig. 6. Multiplicity distributions for the shower energies of 40 and 80 GeV fitted to formula (6).

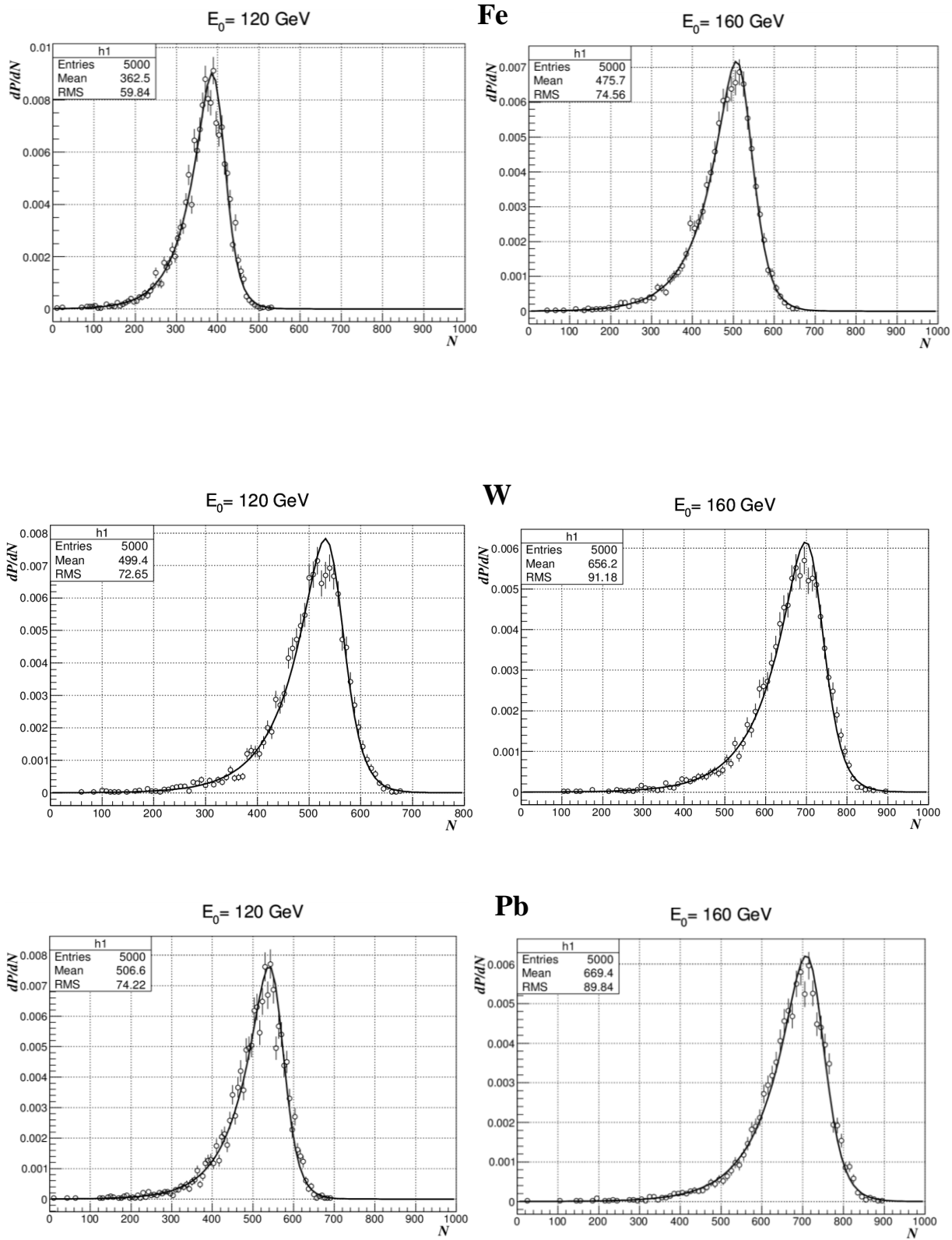

Fig. 7. Multiplicity distributions for the shower energies of 120 and 160 GeV fitted to formula (6).

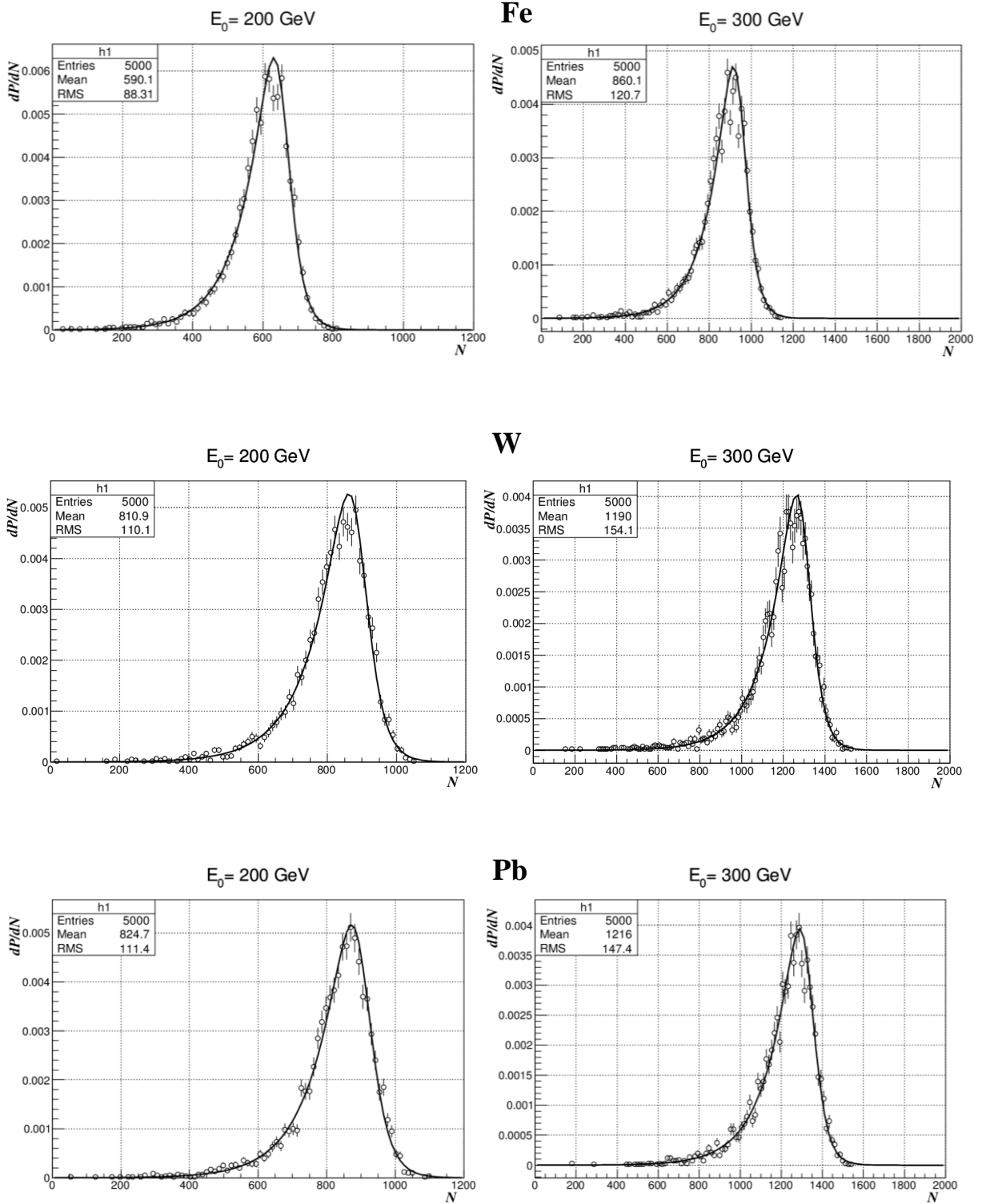

Fig. 8. Multiplicity distributions for the shower energies of 200 and 300 GeV fitted to formula (6).

**Fe**

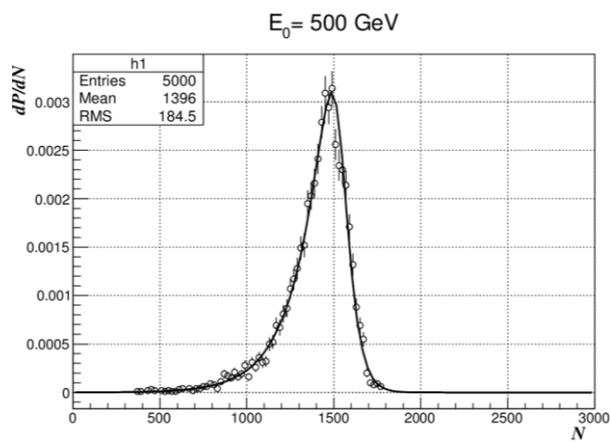
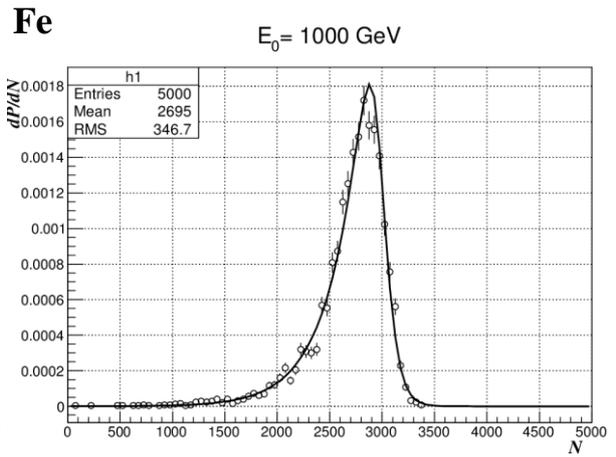

**W**

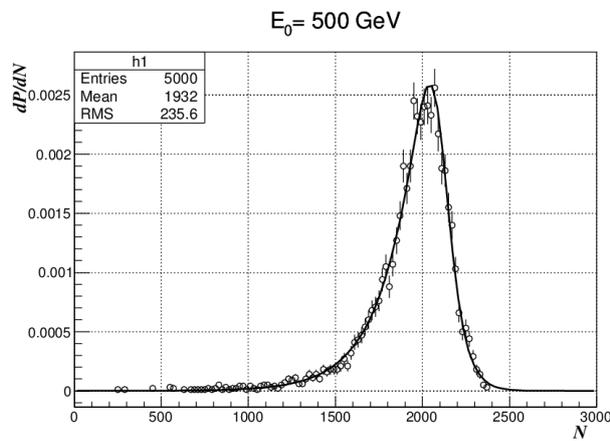
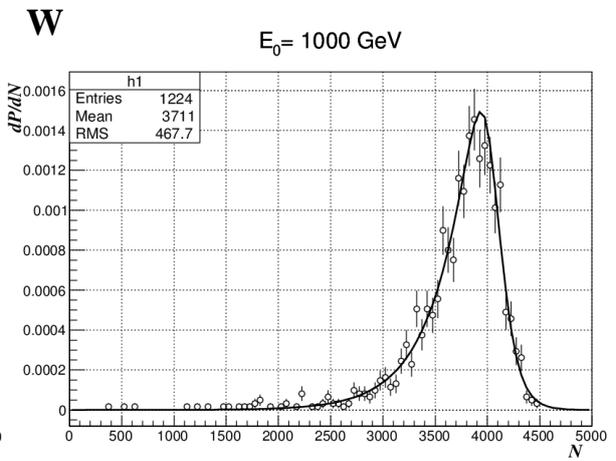

**Pb**

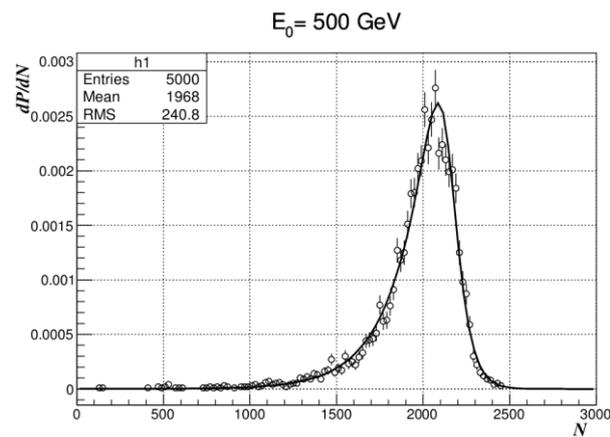
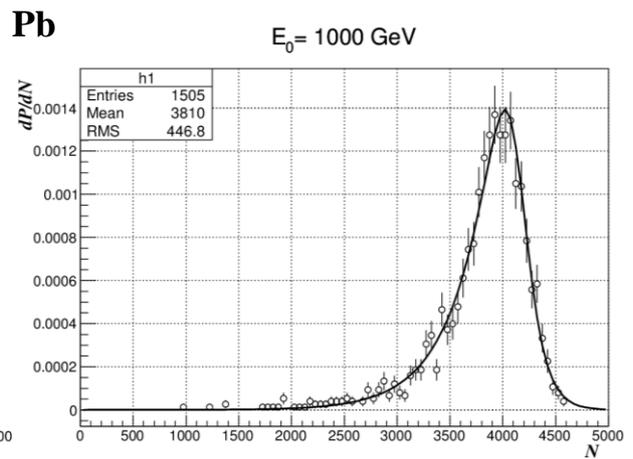

Fig. 9. Multiplicity distributions for the shower energies of 500 and 1000 GeV fitted to formula (6).

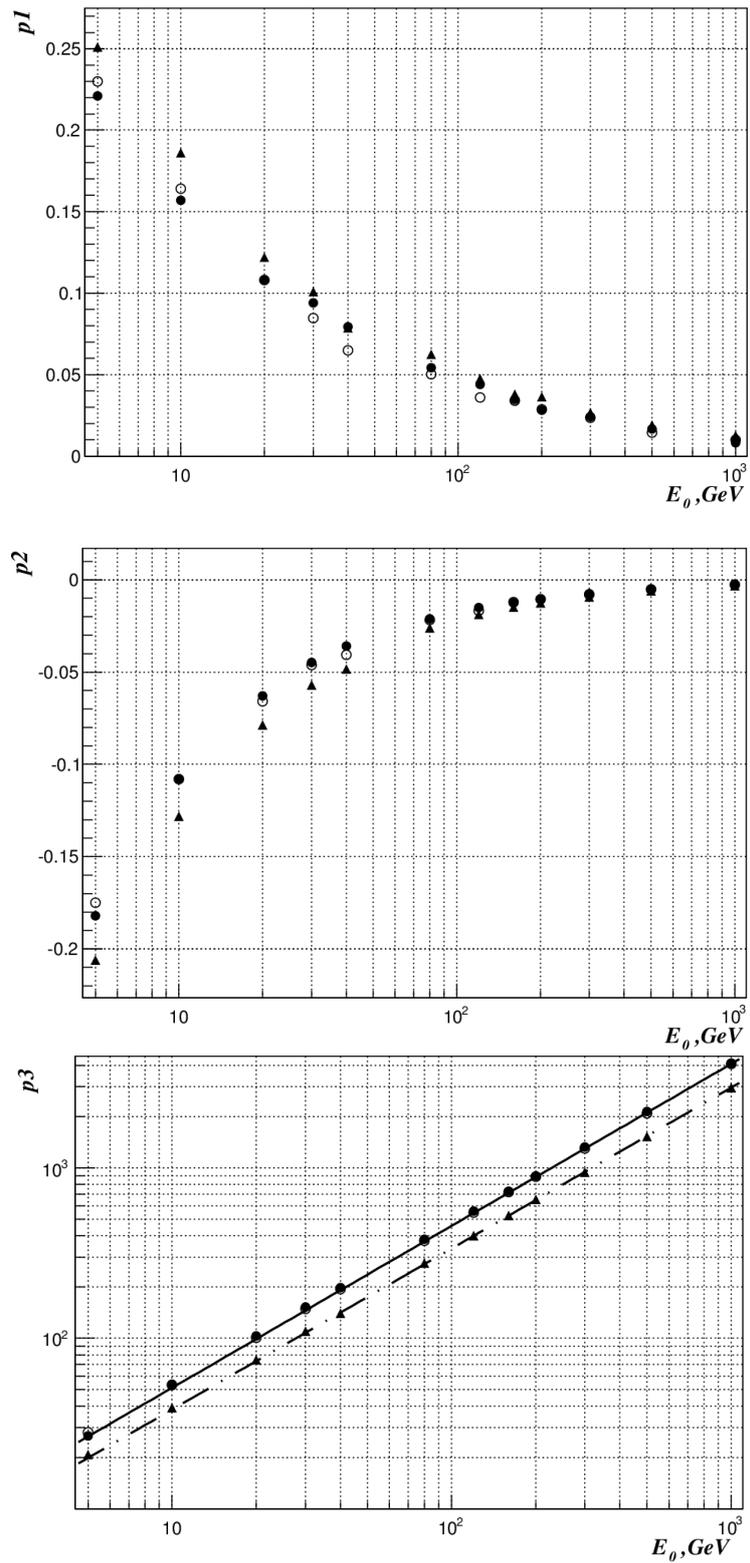

Figure 10. $p_1$, $p_2$ and $p_3$ parameters vs $E_0$ for Fe (▲), W (○) and Pb (●). The solid and dash-dotted lines are the fits to formula (9) for W and Fe with parameters shown in the Table 4. The $p_3$ values for W and Pb almost coincide.

Table 5. $p_{1,2,3}$ parameters for Fe.

| $E_0$, GeV | $p_1$ | $p_2$ | $p_3$ |
|---|---|---|---|
| 5 | 0.251±0.005 | -0.206±0.005 | 21.1±0.2 |
| 10 | 0.186±0.004 | -0.128±0.003 | 39.8±0.3 |
| 20 | 0.122±0.005 | -0.079±0.003 | 74.9±0.9 |
| 30 | 0.101±0.003 | -0.057±0.001 | 110.3±0.7 |
| 40 | 0.0787±0.0022 | -0.0484±0.0011 | 140.6±0.8 |
| 80 | 0.0625±0.0019 | -0.0261±0.0005 | 277.0±1.2 |
| 120 | 0.0476±0.0013 | -0.0187±0.0004 | 400.9±1.5 |
| 160 | 0.0379±0.0012 | -0.0149±0.0003 | 525.2±1.9 |
| 200 | 0.0364±0.0011 | -0.0124±0.0003 | 653.3±2.1 |
| 300 | 0.0266±0.0008 | -0.0094±0.0002 | 944.4±2.9 |
| 500 | 0.0190±0.0007 | -0.0059±0.0001 | 1535±4 |
| 1000 | 0.0126±0.0004 | -0.0032±0.0001 | 2969±6 |

Table 6. $p_{1,2,3}$ parameters for W.

| $E_0$, GeV | $p_1$ | $p_2$ | $p_3$ |
|---|---|---|---|
| 5 | 0.230±0.004 | -0.175±0.004 | 28.0±0.2 |
| 10 | 0.164±0.003 | -0.1082±0.0019 | 53.3±0.3 |
| 20 | 0.111±0.005 | -0.067±0.003 | 100.8±0.9 |
| 30 | 0.086±0.003 | -0.046±0.001 | 148.7±0.8 |
| 40 | 0.076±0.002 | -0.039±0.001 | 194.2±0.9 |
| 80 | 0.055±0.002 | -0.0213±0.0005 | 374.4±1.3 |
| 120 | 0.043±0.001 | -0.0158±0.0003 | 548.7±1.7 |
| 160 | 0.035±0.001 | -0.0122±0.0002 | 720.9±2.2 |
| 200 | 0.0301±0.0009 | -0.0104±0.0002 | 888.2±2.5 |
| 300 | 0.0238±0.0008 | -0.0078±0.0002 | 1299±3.2 |
| 500 | 0.0151±0.0005 | -0.0051±0.0001 | 2096±5.1 |
| 1000 | 0.0092±0.0007 | -0.0028±0.0001 | 4035±17.8 |

Table 7. $p_{1,2,3}$ parameters for Pb.

| $E_0$, GeV | $p_1$ | $p_2$ | $p_3$ |
|---|---|---|---|
| 5 | 0.221±0.004 | -0.183±0.004 | 28.1±0.2 |
| 10 | 0.157±0.003 | -0.1084±0.0020 | 53.6±0.3 |
| 20 | 0.108±0.005 | -0.063±0.002 | 103.2±1.0 |
| 30 | 0.094±0.003 | -0.045±0.001 | 152.6±0.7 |
| 40 | 0.079±0.002 | -0.0360±0.0008 | 198.8±0.9 |
| 80 | 0.054±0.002 | -0.0211±0.0004 | 382.2±1.3 |
| 120 | 0.044±0.001 | -0.0150±0.0003 | 558.4±1.7 |
| 160 | 0.035±0.001 | -0.0125±0.0002 | 730.5±2.1 |
| 200 | 0.0282±0.0009 | -0.0105±0.0002 | 898.5±2.5 |
| 300 | 0.0242±0.0007 | -0.0075±0.0002 | 1327±3 |
| 500 | 0.0165±0.0006 | -0.0049±0.0001 | 2144±5 |
| 1000 | 0.0081±0.0006 | -0.00273±0.00014 | 4123±20 |

## 4. Conclusions

Simulations of EM showers initiated by 5 to 1000 GeV electrons in Fe, W, and Pb are performed using GEANT4. Studies of charge particles multiplicity distributions at the shower maximum show that they are reasonably well described by the inverse sum of two exponents with three free parameters for all materials and energies and the energy dependence of the average multiplicity follows power law with the power of ~0.95. Data presented in the Tables 4-8 and Fig. 10 allow to calculate the multiplicity distribution at any energy within the studied interval.

## Acknowledgment

We gratefully acknowledge the help of D.S. Denisov, T.Z. Gurova, A.V.Popov and D.A. Stoyanova in preparation of this manuscript. This work was supported by the Russian Foundation for Basic Research, project #17-02-00120.